\documentclass[
]{ceurart}

\usepackage{listings}
\begin{document}

\copyrightyear{2024}
\copyrightclause{Copyright for this paper by its authors.
  Use permitted under Creative Commons License Attribution 4.0
  International (CC BY 4.0).}

\conference{CHIRP 2024: Transforming HCI Research in the Philippines Workshop, May 09, 2024, Binan, Laguna}

\title{That Flick is Sick: Gyroscope Integration in Xbox Controllers}

\author[]{Jhervey Edric Cheng}[%
email=jhervey_edric_cheng@dlsu.edu.ph,%
]
\author[]{Stacy Selena Kalaw}[%
email=stacy_selena_kalaw@dlsu.edu.ph,%
]
\author[]{James Patrick Kok}[%
email=james_patrick_kok@dlsu.edu.ph,%
]
\author[]{Alyssa Ysabelle Meneses}[%
email=alyssa_meneses@dlsu.edu.ph,%
]
\author[]{Richard Sy}[%
email=richard_t_sy@dlsu.edu.ph,%
]
\author[]{Jordan Aiko Deja}[%
email=jordan.deja@dlsu.edu.ph,%
orcid=0001-9341-6088%
]

\address[1]{De La Salle University, Manila, Philippines}


\begin{abstract}
 Gyroscope integration in Xbox controllers offers new possibilities for enhancing gaming experiences, particularly in first-person shooter (FPS) games. To investigate its potential, we conducted an empirical study with 11 participants, comparing aim precision and reaction times across three input methods: a computer mouse, a standard Xbox controller, and a gyroscope-enabled controller. Participants completed an aim training task, revealing the mouse as the most accurate device, followed by the standard controller. Interestingly, the gyroscope-enabled controller showed reduced accuracy and slower reaction times, attributed to challenges in sensitivity and control. Participant feedback highlighted areas for improvement, including refined sensitivity settings, control stability, and software design. These findings underscore the need for design innovations, such as camera rotation limits and optimized sensitivity thresholds, to make gyroscope-enabled controllers more competitive. Future work should consider diverse gamer profiles and extended evaluation contexts to better understand the role of gyroscopes in gaming interfaces.
\end{abstract}


\maketitle

\section{Introduction}
\par Technological advancements in gaming continue to reshape how players engage with virtual worlds. As gaming environments become increasingly dynamic and immersive, new challenges emerge, particularly in first-person shooter (FPS) games, where precise aiming is paramount. Enhancing input devices is a critical avenue for addressing these challenges. One promising approach involves integrating gyroscopes into Xbox controllers, offering potential improvements in control and responsiveness for FPS players. Gyroscopes, which consist of a spinning wheel, disk, or circulating beam of light capable of rotating about an axis in any direction \cite{1}, are inertial sensors that can measure or control orientation and rotational velocity \cite{6}. By embedding gyroscopes into controllers, gaming interfaces could unlock novel motion-based interaction techniques, enriching player experiences.

\par A gyroscope-integrated Xbox controller could bridge the gap between traditional input methods and motion-based gestures, enabling players to interact with virtual environments more intuitively \cite{3}. This hybrid input model promises to enhance control, responsiveness, and immersion, particularly in FPS games. Previous research has explored gyroscope applications in gaming, but their integration into mainstream controllers remains relatively novel. For example, \cite{Gerling2011} noted usability challenges when introducing new input devices in FPS games. However, these challenges diminished when games were well-designed, leading to positive player experiences.

\par Other studies have evaluated gyroscope functionality in different contexts. For instance, researchers examined the built-in gyrosensor in the Steam Controller for pointing tasks \cite{4}, but its implementation mimicked a steering wheel rather than leveraging tilt-based inputs. Moreover, their study was conducted in a 2D environment, limiting insights into 3D gaming scenarios. Another study directly investigated gyroscope use in FPS games, revealing that while it enhanced realism, performance with traditional mouse and keyboard inputs remained superior \cite{5}. This outcome was attributed to user familiarity with conventional input methods and variations in gyroscope sensitivity settings, as sensitivity coefficients significantly affect performance \cite{7}.

\par The present study seeks to build on these findings through a comprehensive experiment comparing three input devices: a computer mouse, a standard Xbox controller, and a gyroscope-enabled Xbox controller. Eleven participants engaged in an aim training task, allowing for the empirical evaluation of aim precision and reaction times. The study also investigates potential implications, such as the learning curve associated with gyroscope use and the importance of sensitivity calibration in optimizing gameplay performance.

\par Sensitivity settings, in particular, play a crucial role in gaming performance. For example, \cite{Boudaoud2023} demonstrated that adjusting sensitivity levels affects in-game outcomes: lower sensitivities slow rotational speeds, while higher sensitivities reduce motion precision, requiring more corrective actions and increasing task completion time. By examining sensitivity's impact, this study aims to identify calibration strategies that balance precision and responsiveness in gyroscope-based gaming.

\par Ultimately, this research aims to inform the design of next-generation gaming interfaces by evaluating gyroscope integration's effects on control, aiming accuracy, reaction times, and overall player experience. Findings will include both quantitative data and participant feedback, offering insights into the intuitiveness and transformative potential of gyroscope-enabled controllers in FPS games.

\section{Method}
\par To explore the integration of gyroscope technology into gaming, the MPU6050 Accelerometer and Gyroscope sensor was employed. This sensor was mounted on an Arduino board and affixed to the topmost center of an Xbox controller, ensuring stability and ease of use during gameplay. To provide a controlled environment for evaluation, the researchers developed a custom JavaScript (JS) first-person shooter (FPS) game specifically designed for this experiment. Figure 1 illustrates the experimental setup, showcasing the game running on a laptop and the modified Xbox controller equipped with the gyroscope sensor.

\subsection{Experiment Procedure}

\begin{figure}[htb]
  \centering
  \includegraphics[width=0.5\linewidth]{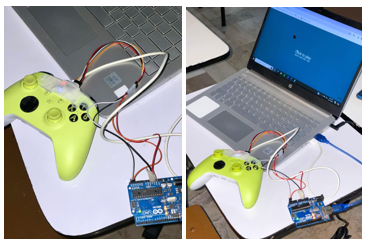}
  \caption{Experimental Setup}
  \label{fig:(1)}
\end{figure}

\par This study evaluated aim precision in a first-person shooter (FPS) game using three input devices: a regular computer mouse, an Xbox controller without a gyroscope, and an Xbox controller integrated with a gyroscope. The experiment aimed to investigate how different input devices influence user performance and experience in gaming tasks \cite{Hasanain2024}.

\par Eleven participants were recruited for the study, all seated comfortably in office chairs throughout the experiment to ensure consistent posture. Before gameplay, participants were given a brief introduction to the controls for each input device and the game mechanics. To minimize performance variability, participants were instructed to maintain the same posture and grip on the input device throughout the session. Proper seating and posture were emphasized to enhance performance and reduce discomfort, aligning with findings by \cite{Martha2022}, which highlight the importance of ergonomic arrangements during prolonged gaming sessions.

\par Each participant completed three rounds of gameplay, one for each input device, with short breaks between rounds to prevent fatigue. During each round, players engaged in a shooting simulation task, aiming to hit 20 targets. These targets spawned randomly across 10 fixed positions and moved horizontally within a predefined range before reversing direction.

\par To ensure consistency in the experiment, input device settings, such as sensitivity, were standardized across all participants and configurations. Participants were not allowed to adjust these preferences, maintaining uniform conditions for evaluating device performance.

\subsection{Data Collection}
\par The game recorded players’ accuracy and response times across the three input devices. Each round's total time was logged as the overall reaction time. Accuracy was determined by calculating the ratio of targets hit to total shots fired by each player. These metrics were automatically exported into a spreadsheet file for subsequent analysis.

\par Following the gameplay sessions, participants were invited to provide feedback and share insights about their experiences with the various input devices. This qualitative feedback offered valuable perspectives on the usability, intuitiveness, and challenges of using the gyroscope-enabled controller. These reflections enabled the researchers to better understand user perceptions and attitudes toward adopting new input technologies in gaming contexts.



\section{Results and Analysis}
\subsection{Mouse Input}
\par When using the mouse, participants achieved an average accuracy of 53.43\%, with notable variability, as reflected by a standard deviation of 25.24. Accuracy scores ranged from a high of 86\% to a low of 16\%, indicating significant differences in individual abilities, skills, and comfort levels with the mouse as an input device.

\par In terms of time, participants completed the task in an average of 122 seconds, corresponding to an average reaction time of 6.11 seconds per target. Figure 2 illustrates the distribution of accuracy and average reaction times among participants. As shown, accuracy scores vary widely, and two outliers are observed in the reaction times. These outliers could result from individual differences or external factors, such as the constraints imposed by the study.

\par One important consideration is that while participants were likely familiar with using a mouse, the inability to calibrate sensitivity during the experiment may have introduced difficulties. As \cite{Boudaoud2023} highlights, sensitivity adjustments play a critical role in improving performance, and their absence might have contributed to the performance inconsistencies and outliers observed in this study.

\begin{figure}[htb]
  \centering
  \includegraphics[width=0.58\linewidth]{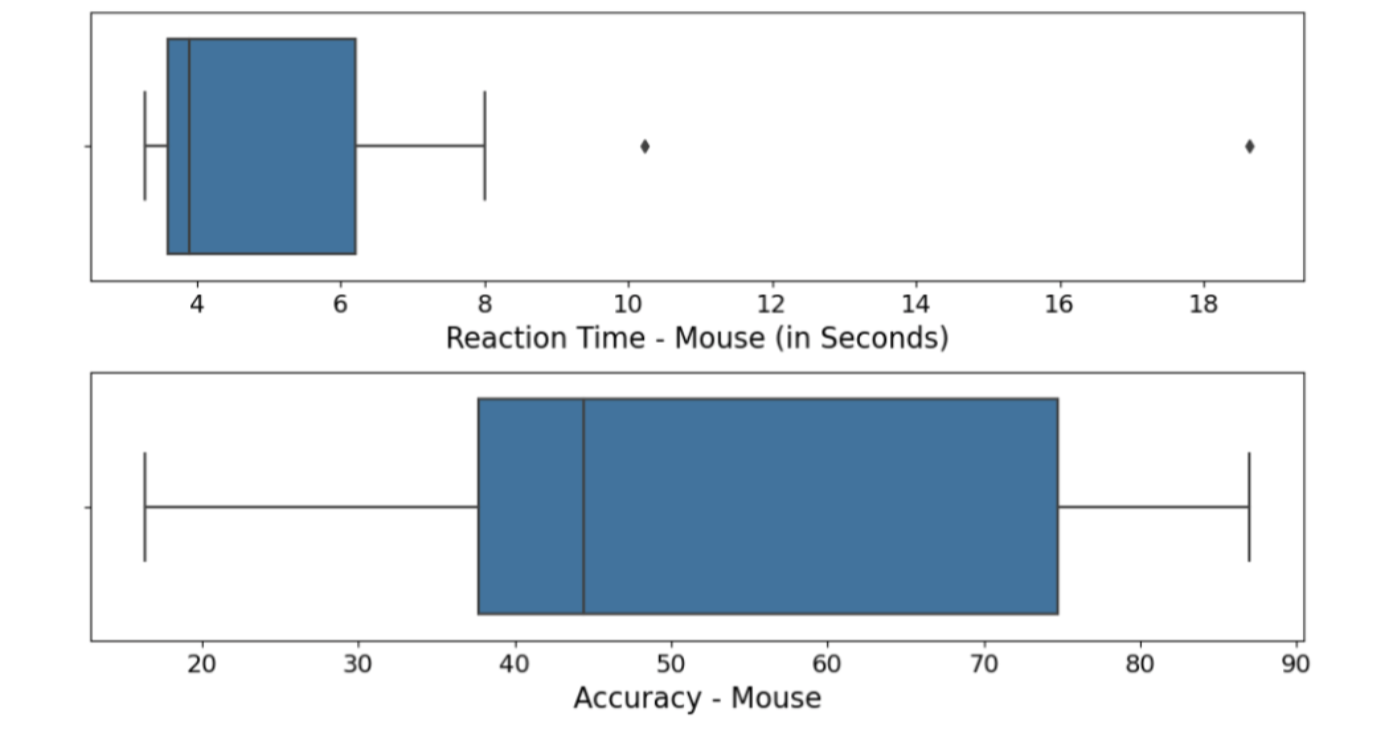}
  \caption{Boxplot of Data Using Mouse as Input}
  \label{fig:(2)}
\end{figure}

\subsection{Controller Input}
\par The second part of the analysis explores the use of the Xbox controller without a gyroscope as an input device. Accuracy and reaction time measurements reveal detailed trends in player behavior, which are crucial for evaluating the effectiveness of this input method. The average accuracy of participants was 41.51\%, with a standard deviation of 9.7, reflecting relatively low variability compared to the mouse input. The average total response time was 134.66 seconds, resulting in an average of 6.73 seconds per target hit.

\par The lower variation in accuracy with this input device could be attributed to participants’ familiarity with using the controller. Unlike the mouse, where participants may have varying levels of experience, most participants seemed to be more comfortable with the controller for aiming tasks. Figure 3 illustrates the distribution of accuracy and reaction times. As seen in the figure, the data for the controller input exhibits less variability and no outliers. Furthermore, all reaction times are single-digit values, suggesting a more consistent performance across participants. While the average accuracy for the controller is similar to that of the mouse, the range of accuracy is narrower, indicating more uniform performance across the group.


\begin{figure}[htb]
  \centering
  \includegraphics[width=0.6\linewidth]{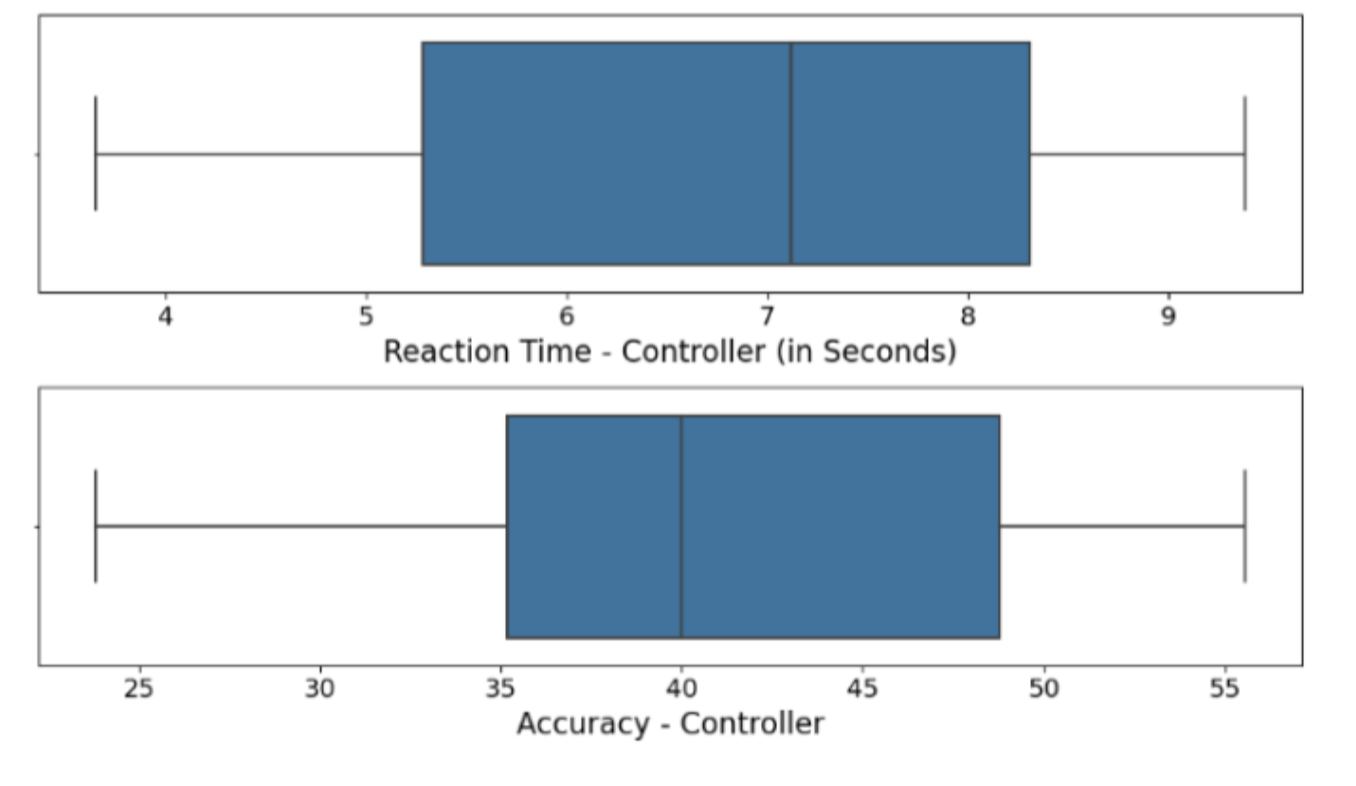}
  \caption{Boxplot of Data Using Controller as Input}
  \label{fig:(3)}
\end{figure}

\subsection{Gyroscope-integrated controller input}
\par Participants demonstrated similar performance with the gyroscope-augmented controller as with the original controller, achieving an average accuracy of 33.52\% with a standard deviation of 8.23. The average total response time was 245.16 seconds, corresponding to an average of 12.26 seconds per target hit.

\par Figure 4 shows that the reaction time data for the gyroscope controller exhibits greater variation compared to the mouse and original controller inputs. Additionally, the reaction times for the gyroscope controller are relatively larger than those observed for the other two devices. In contrast to the mouse and controller inputs, the accuracy with the gyroscope controller does not exceed 50

\begin{figure}[htb]
  \centering
  \includegraphics[width=0.6\linewidth]{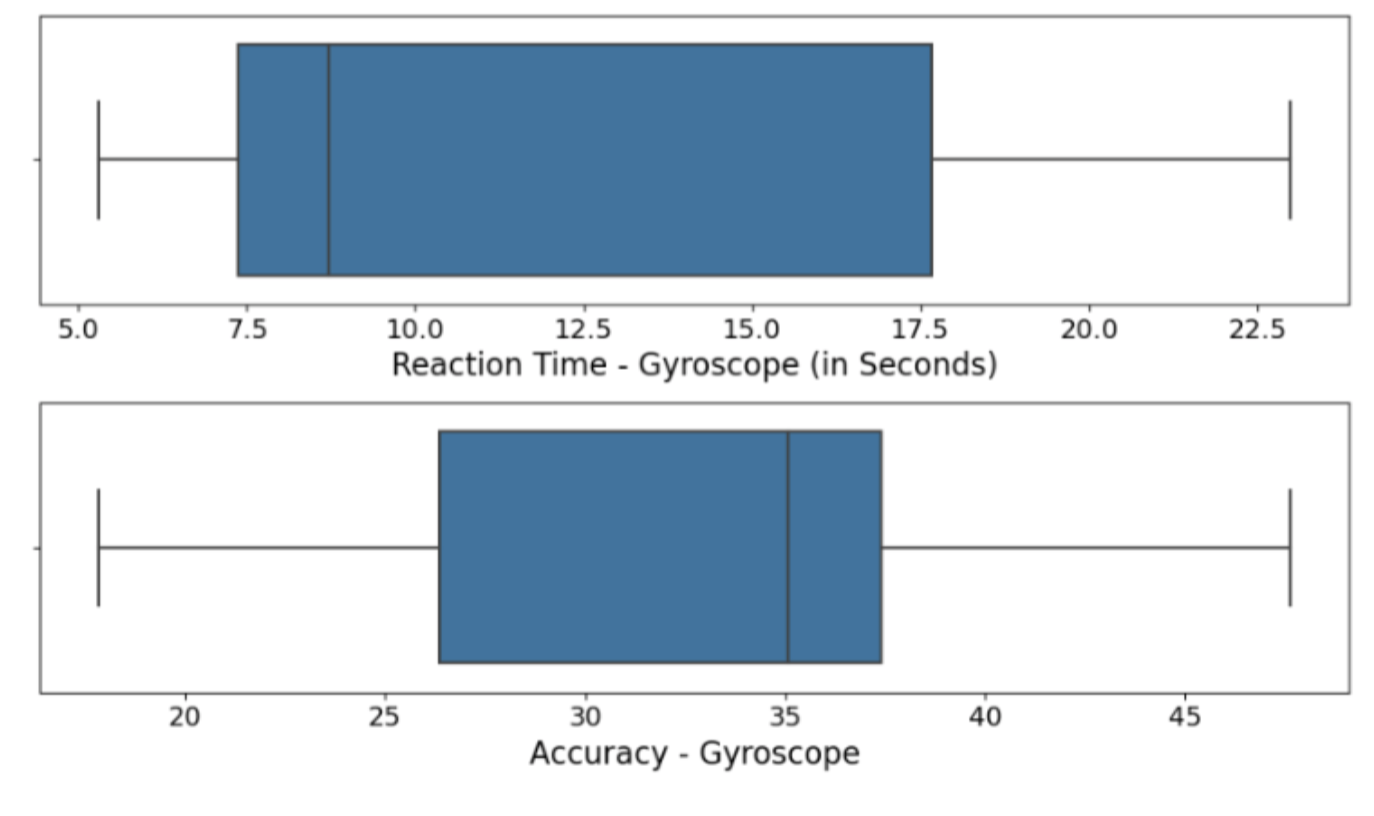}
  \caption{Boxplot of Data Using Gyroscope as Input}
  \label{fig:(4)}
\end{figure}

\subsection{Comparison among inputs}
\par This section compares the accuracy across the different input devices. Notably, players using the mouse achieved the highest average accuracy of approximately 53.43\%, followed by those using a controller without a gyroscope, with an average of 41.51\%. Players using a controller with a gyroscope had the lowest average accuracy, around 33.52\%. This disparity suggests that the mouse provides users with a more precise targeting mechanism compared to controllers, likely due to the fine-grained and direct control offered by mouse movements.

\par Figure 5 displays the accuracy values of each participant across the different input devices. Notably, Participants 1 and 4 demonstrated lower accuracy with the mouse compared to the other two input devices, which contrasts with the overall trend in average values. Additionally, some participants, such as Participants 2 and 3, exhibited similar accuracy values across all input devices, suggesting they were able to adapt easily to the differences between the devices. Finally, Participants 7, 8, and 10 showed particularly high accuracy with the mouse but lower accuracy with the other input methods, indicating a potential preference or proficiency with the mouse for precision tasks.


\begin{figure}[htb]
  \centering
  \includegraphics[width=0.6\linewidth]{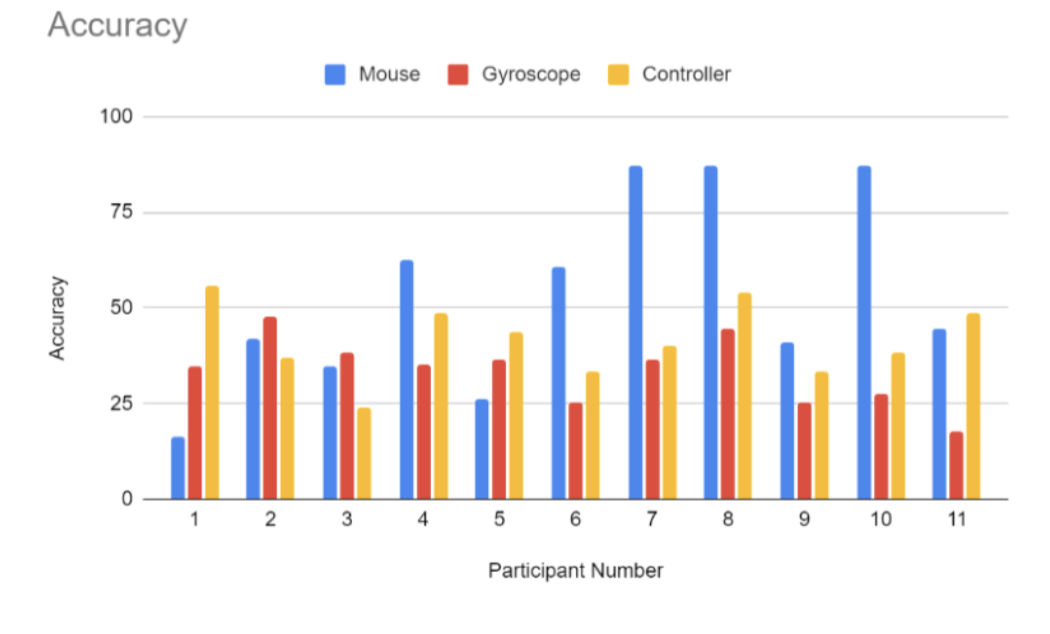}
  \caption{Accuracy of the Participants Across Different Inputs}
  \label{fig:(5)}
\end{figure}

\par In terms of reaction time, Figure 6 shows the participants' average reaction times across the different input devices. Notably, Participants 4, 5, and 7 exhibited relatively higher reaction times with the gyroscope compared to the other participants. Furthermore, only Participant 5 had a high reaction time with the mouse as well. These observations suggest that these participants may have faced challenges in adjusting to the different input devices.

\par Across all participants, the controller demonstrated the most consistent reaction times. This consistency may be attributed to the more limited range of movement offered by the controller joystick, which restricts movement to how far the joystick can travel. It is also noteworthy that Participants 2, 3, 6, 8, 9, 10, and 11 showed relatively similar reaction times with the gyroscope as with the other input devices.

\par Overall, the reaction time with the gyroscope was, on average, larger than with the other input devices. This difference can likely be attributed to several factors, such as participants' familiarity and biases toward existing input devices and the difficulty they experienced in adjusting to learning a new input device \cite{8}.


\begin{figure}[htb]
  \centering
  \includegraphics[width=0.6\linewidth]{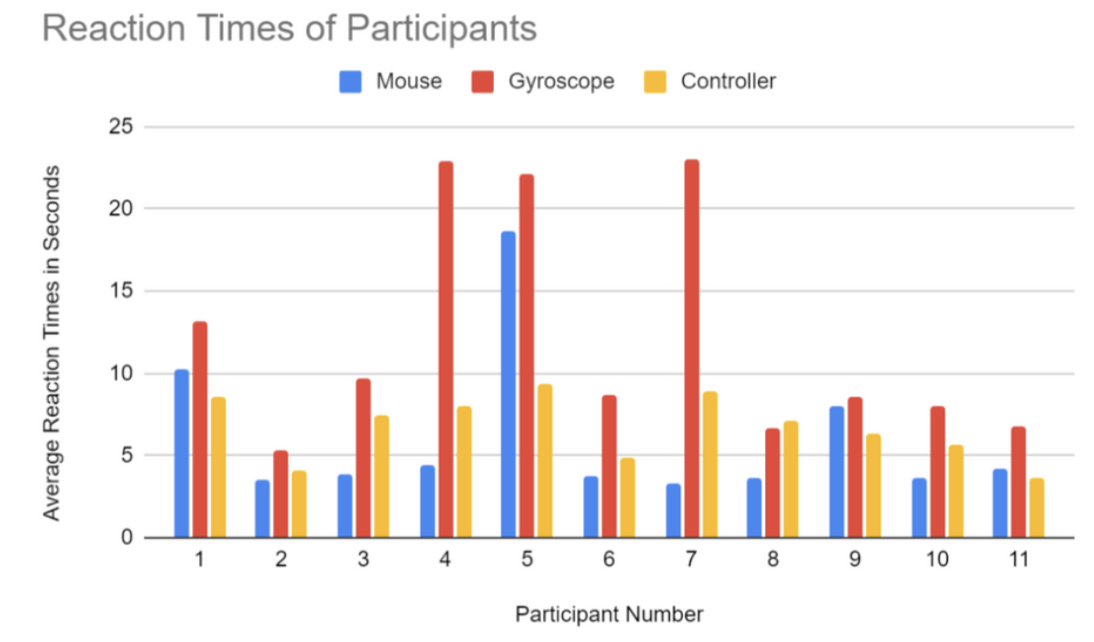}
  \caption{Reaction Times of the Participants Across Different Inputs}
  \label{fig:(6)}
\end{figure}

\par Figure 7 presents a plot of the average reaction time (in seconds) of participants against their accuracy for each input device. In theory, higher reaction times may suggest higher accuracy, as users would take more time to aim precisely before shooting. However, the figure reveals instances where participants achieved high accuracy with low reaction times and, conversely, low accuracy with high reaction times. Notably, the data points with high accuracy and low reaction time correspond to the mouse input, while those with low accuracy and high reaction time are associated with the gyroscope input. This observation is significant, as it may suggest that participants found the gyroscope more difficult to use and, conversely, felt more comfortable with the mouse, which might have contributed to better performance and quicker reactions.

\begin{figure}[htb]
  \centering
  \includegraphics[width=0.55\linewidth]{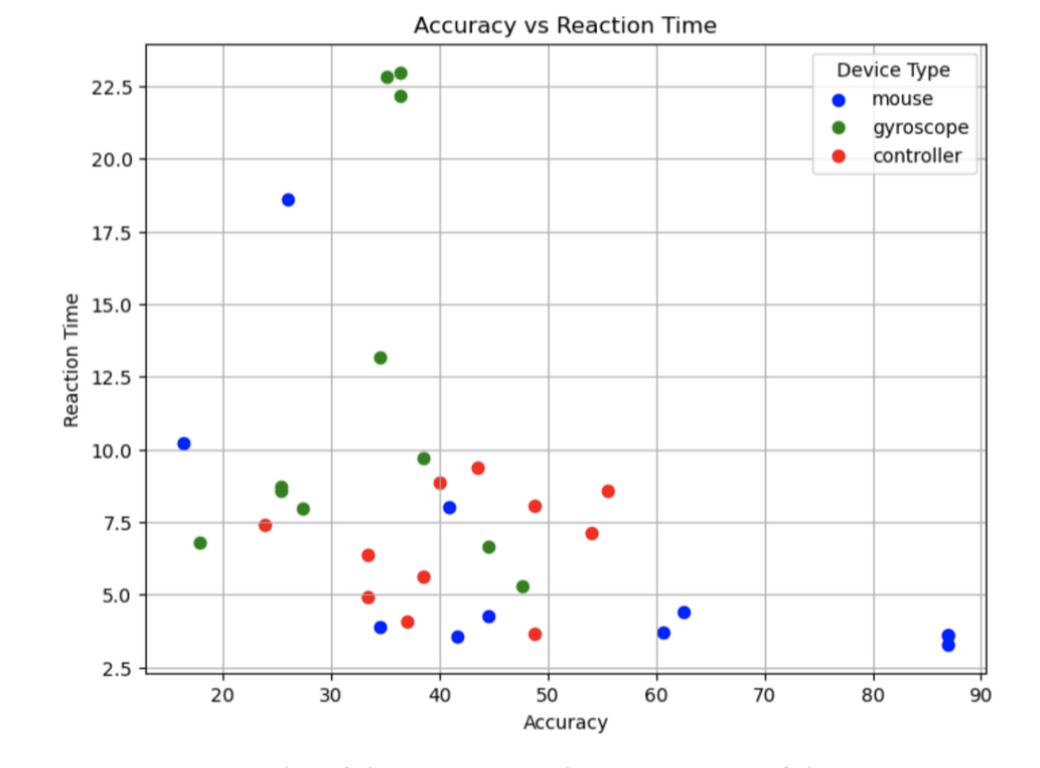}
  \caption{Scatter Plot of the Accuracy and Reaction Time of the Participants}
  \label{fig:(7)}
\end{figure}

\par To confirm the observations made from the earlier figures, a hypothesis test was conducted. Given the small sample size of only 11 data points, a t-test was performed to compare the reaction times across the three input devices. The test was conducted with an alpha level of 0.05. Snippet 1 presents the results of the hypothesis testing between the input devices. As shown in the snippet, there is no significant difference in reaction times between the mouse and the controller. However, there is a significant difference between both the mouse and controller compared to the gyroscope. These results further support the initial observations that using the controller with the gyroscope may lead to higher reaction times compared to the other two input devices.

\begin{center}
    \begin{minipage}{1\linewidth}
    \centering
    \includegraphics[width=0.6\linewidth]{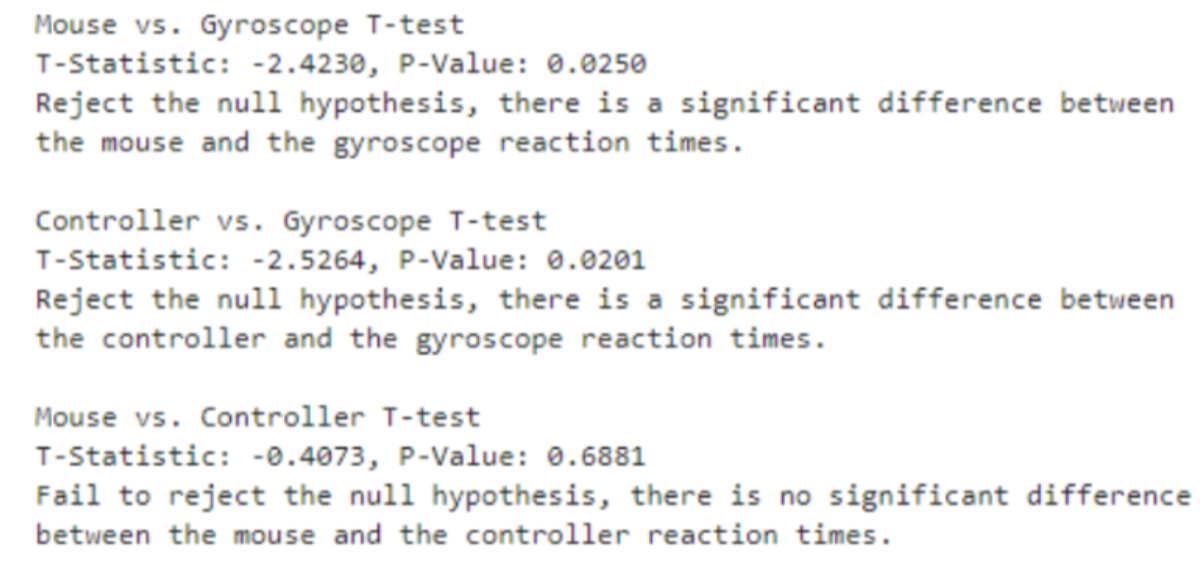}
    \end{minipage}
    \begin{minipage}{1\linewidth}
    \centering
    \textbf{Snippet 1: T-test Results} \\[0.5em]
    \end{minipage}
\end{center}

\subsection{Qualitative Data Analysis}
\par In addition to the quantitative data, participants provided valuable feedback regarding their experiences with the various input devices. One participant noted that the gyroscope feature was highly sensitive to movement, which often resulted in overshooting targets—an issue echoed by others who found it challenging to maintain control due to the constant camera movement. Several participants also expressed concerns about the length of the controller's cord, predicting that it could negatively affect the overall gaming experience.

\par On a more positive note, many participants praised the gyroscopic feature, mentioning that it added a sense of realism to the shooting mechanics by aligning with their hand movements. However, opinions on sensitivity were mixed: some participants found the gyroscope less responsive, while others felt that it was off-center, which led to a disconnected gameplay experience. Furthermore, some participants cited software-related issues, such as the lack of depth and realism in the background, which made the task more difficult.

\par When asked about their preferences, one participant favored the controller without the gyroscope, while three preferred the gyroscope-augmented controller. However, the majority of participants favored the mouse as their preferred input device for aiming. These varying opinions highlight the importance of considering user feedback when designing and refining game interfaces. They also suggest areas for potential improvements, both in the hardware and software components of the experiment.

\section{Discussion and Design Implications}
\subsection{Software improvements}
\par The findings of this study suggest several avenues for improvement in the design of gyroscope-augmented controllers. One key concern expressed by participants was the unintended camera movements, which were often caused by the sensitivity of camera rotation. To address this, one potential improvement could involve imposing limits on camera rotation. For example, players could be restricted to rotating the camera only up to 90 degrees along both the positive and negative axes, providing a more controlled and predictable experience.

\par Additionally, enhancing the game environment’s visual design could improve the user's ability to perceive the three-dimensional aspects of the space. Currently, the game uses a single color for the floor, walls, and ceiling, which may hinder depth perception. A diversified color palette, along with adjustments to the field of view (FOV) and target sizes, could make the environment more immersive and easier to navigate. Incorporating enhanced graphics and varied color schemes would further boost user engagement.

\par Moreover, conducting usability testing could help identify and mitigate any confounding variables that might impact the user experience. This observation aligns with previous studies highlighting the importance of realism in gyroscope-based input devices \cite{5}. In this case, improving the software’s realism is essential to fully leverage the gyroscope's potential for enhancing gameplay immersion.


\subsection{Sensitivity adjustments}
\par Some participants also suggested that the sensitivity of the gyroscope be addressed. One potential solution could involve implementing a threshold mechanism, where camera movement is only registered when the change in gyroscope position exceeds a certain threshold value. This could help to mitigate unintended camera shifts and improve user control.

In addition, sensitivity remains an area for exploration in future studies. Theoretically, accuracy may be higher with lower sensitivity, as it allows players to be more precise and deliberate with their aim. Conversely, higher sensitivity could lead to lower accuracy, as it increases the likelihood of over- and under-shooting. Generally, games offer a wide range of sensitivity settings to accommodate a variety of players, as sensitivity preferences are influenced by factors such as physical space, grip on the input device, and individual preferences. Nonetheless, it would be interesting to determine the sensitivity level at which most users find the input device easiest to control, and how this affects their overall performance and comfort.

\subsection{Recommendations for Future Work}
\par Expanding the scope of the current study presents several promising avenues for future research. Increasing the participant pool to include a more diverse demographic could provide valuable insights into the broader applicability and effectiveness of gyroscope integration in gaming controllers. A larger and more varied sample may yield more comprehensive insights into how different groups of users interact with gyroscope-augmented controllers. Additionally, incorporating participant background analysis, such as input device preferences and gaming experience, could offer deeper understanding of how these factors influence the adoption and performance with gyroscopic controls.

\section{Conclusion}
\par This study explores the impact of gyroscope integration in Xbox controllers on gaming performance. While the mouse remains the preferred input device for the aim training game, the gyroscope introduces challenges related to sensitivity and control. To improve user experience, design enhancements such as limiting camera rotation and refining graphics could be beneficial. Future research should expand the participant pool and incorporate background analysis, such as gaming experience and input preferences, to better assess the effectiveness and applicability of gyroscope technology in gaming controllers. By addressing user feedback and refining both hardware and software design, future studies can drive the development of more immersive and responsive gaming interfaces.
\bibliography{main}

\begin{thebibliography}{11}
\expandafter\ifx\csname natexlab\endcsname\relax\def\natexlab#1{#1}\fi
\providecommand{\url}[1]{\texttt{#1}}
\providecommand{\href}[2]{#2}
\providecommand{\path}[1]{#1}
\providecommand{\DOIprefix}{doi:}
\providecommand{\ArXivprefix}{arXiv:}
\providecommand{\URLprefix}{URL: }
\providecommand{\Pubmedprefix}{pmid:}
\providecommand{\doi}[1]{\href{http://dx.doi.org/#1}{\path{#1}}}
\providecommand{\Pubmed}[1]{\href{pmid:#1}{\path{#1}}}
\providecommand{\bibinfo}[2]{#2}
\ifx\xfnm\relax \def\xfnm[#1]{\unskip,\space#1}\fi
\bibitem[{YoungWonks(2021)}]{1}
\bibinfo{author}{YoungWonks}, \bibinfo{title}{What is a gyroscope and how does it work?}, \bibinfo{year}{2021}. \URLprefix \url{https://www.youngwonks.com/blog/What-is-a-Gyroscope-and-How-Does-It-Work}.
\bibitem[{Gill et~al.(2022)Gill, Howard, Mazhar, and McKee}]{6}
\bibinfo{author}{W.~A. Gill}, \bibinfo{author}{I.~Howard}, \bibinfo{author}{I.~Mazhar}, \bibinfo{author}{K.~McKee},
\newblock \bibinfo{title}{A review of mems vibrating gyroscopes and their reliability issues in harsh environments},
\newblock \bibinfo{journal}{Sensors} \bibinfo{volume}{22} (\bibinfo{year}{2022}). \URLprefix \url{https://www.mdpi.com/1424-8220/22/19/7405}. \DOIprefix\doi{10.3390/s22197405}.
\bibitem[{Haris et~al.(2021)}]{3}
\bibinfo{author}{D.~Haris}, et~al., \bibinfo{title}{Gyroscope implementation in arcade shooter game "full metal parabellum"}, \bibinfo{year}{2021}. \URLprefix \url{https://lintar.untar.ac.id/repository/penelitian/buktipenelitian_10809003_4A020322105100.pdf}.
\bibitem[{Gerling et~al.(2011)Gerling, Klauser, and Niesenhaus}]{Gerling2011}
\bibinfo{author}{K.~Gerling}, \bibinfo{author}{M.~Klauser}, \bibinfo{author}{J.~Niesenhaus},
\newblock \bibinfo{title}{Measuring the impact of game controllers on player experience in fps games}  (\bibinfo{year}{2011}). \DOIprefix\doi{10.1145/2181037.2181052}.
\bibitem[{Ramcharitar and Teather(2017)}]{4}
\bibinfo{author}{A.~Ramcharitar}, \bibinfo{author}{R.~J. Teather},
\newblock \bibinfo{title}{A fitts' law evaluation of video game controllers: thumbstick, touchpad and gyrosensor},
\newblock in: \bibinfo{booktitle}{Proceedings of the 2017 CHI Conference extended abstracts on human factors in computing systems}, \bibinfo{year}{2017}, pp. \bibinfo{pages}{2860--2866}.
\bibitem[{Tokta{\c{s}} and Serif(2019)}]{5}
\bibinfo{author}{A.~O. Tokta{\c{s}}}, \bibinfo{author}{T.~Serif},
\newblock \bibinfo{title}{Evaluation of crosshair-aided gyroscope gamepad controller},
\newblock in: \bibinfo{booktitle}{Mobile Web and Intelligent Information Systems: 16th International Conference, MobiWIS 2019, Istanbul, Turkey, August 26--28, 2019, Proceedings 16}, \bibinfo{organization}{Springer}, \bibinfo{year}{2019}, pp. \bibinfo{pages}{294--307}.
\bibitem[{Krzysztofik and Koruba(2021)}]{7}
\bibinfo{author}{I.~Krzysztofik}, \bibinfo{author}{Z.~Koruba},
\newblock \bibinfo{title}{Study on the sensitivity of a gyroscope system homing a quadcopter onto a moving ground target under the action of external disturbance},
\newblock \bibinfo{journal}{Energies} \bibinfo{volume}{14} (\bibinfo{year}{2021}) \bibinfo{pages}{1696}. \URLprefix \url{https://doi.org/10.3390/en14061696}. \DOIprefix\doi{10.3390/en14061696}.
\bibitem[{Boudaoud et~al.(2023)Boudaoud, Spjut, and Kim}]{Boudaoud2023}
\bibinfo{author}{B.~Boudaoud}, \bibinfo{author}{J.~Spjut}, \bibinfo{author}{J.~Kim},
\newblock \bibinfo{title}{Mouse sensitivity in first-person targeting tasks},
\newblock \bibinfo{journal}{IEEE Transactions on Games} \bibinfo{volume}{15} (\bibinfo{year}{2023}) \bibinfo{pages}{493--506}. \DOIprefix\doi{10.1109/TG.2023.3293692}.
\bibitem[{Hasanain(2024)}]{Hasanain2024}
\bibinfo{author}{B.~Hasanain},
\newblock \bibinfo{title}{The role of ergonomic and human factors in sustainable manufacturing: A review},
\newblock \bibinfo{journal}{Machines} \bibinfo{volume}{12} (\bibinfo{year}{2024}). \URLprefix \url{https://www.mdpi.com/2075-1702/12/3/159}. \DOIprefix\doi{10.3390/machines12030159}.
\bibitem[{Martha et~al.(2022)Martha, Budaya, Agustino, Harsemadi, and Pande}]{Martha2022}
\bibinfo{author}{G.~I.~R. Martha}, \bibinfo{author}{G.~B. A.~G. Budaya}, \bibinfo{author}{D.~P. Agustino}, \bibinfo{author}{G.~Harsemadi}, \bibinfo{author}{M.~S.~A. Pande},
\newblock \bibinfo{title}{An empirical analysis of ergonomic gaming peripherals improving gaming performance},
\newblock \bibinfo{journal}{Journal of Games, Game Art, and Gamification} \bibinfo{volume}{7} (\bibinfo{year}{2022}) \bibinfo{pages}{15–21}. \DOIprefix\doi{10.21512/jggag.v7i1.8258}.
\bibitem[{Swing(2017)}]{8}
\bibinfo{author}{O.~Swing},
\newblock \bibinfo{title}{Using gyroscope technology to implement a leaning technique for game interaction},
\newblock \bibinfo{year}{2017}. \URLprefix \url{https://api.semanticscholar.org/CorpusID:66937523}.

\end{thebibliography}




\end{document}